\documentclass[12pt]{elsarticle}
\biboptions{numbers,sort&compress}

\usepackage{longtable}
\usepackage{amsmath}
\usepackage{amssymb}

\usepackage{enumerate}

\usepackage{setspace}
\usepackage{ltablex}
\usepackage{booktabs}

\usepackage{graphicx}
\usepackage{pdflscape}
\usepackage{xcolor}
\usepackage{lipsum}

\usepackage{accents}

\usepackage{lscape}

\usepackage{ntheorem}
\theoremstyle{plain}
                      {\theorembodyfont{\rmfamily}
                      \theoremseparator{.}
                       
           \newtheorem{thm}{Theorem}[section]
           \theoremstyle{plain}
           
           \theoremstyle{plain} 
           \theoremstyle{plain}

           \theoremstyle{plain}
           \newtheorem{rem}{Remark}[section]}
 \usepackage{makeidx}

\newcounter{rownocnt}
\newcommand{\rowno}{\addtocounter{rownocnt}{1}\arabic{rownocnt}.}

\newcolumntype{C}[1]{>{\centering\arraybackslash}p{#1}}

\newcommand{\classifcase}[1]{}

\usepackage{amssymb,verbatim,amsmath}

\usepackage{hyperref}

\def\dap#1{\mathop {#1}\limits_{+a}}
\def\dam#1{ \mathop{#1}\limits_{-a}}
\def\dtp#1{\mathop {#1}\limits_{+\tau}}
\def\dtm#1{\mathop {#1}\limits_{-\tau}}
\def\dbp#1{\mathop {#1}\limits_{+b}}
\def\dbm#1{\mathop {#1}\limits_{-b}}
\def\dh#1{\mathop {#1}\limits_{h}}

\newcommand{\ppartial}[1]{\frac{\partial}{\partial{#1}}}

\allowdisplaybreaks

\begin{document}

\begin{frontmatter}

\title{Symmetries, conservation laws and difference schemes of
the (1+2)-dimensional shallow water equations in Lagrangian coordinates}

\author[mymainaddress]{V.~A. Dorodnitsyn\corref{mycorrespondingauthor}}
\cortext[mycorrespondingauthor]{Corresponding author}
\ead{Dorodnitsyn@keldysh.ru,dorod2007@gmail.com}

\author[mymainaddress,mysecondaryaddress]{E.~I. Kaptsov}
\ead{evgkaptsov@gmail.com}

\author[mymainaddress,mysecondaryaddress]{S. ~V. Meleshko}
\ead{Meleshko@gmail.com}

\address[mymainaddress]{Keldysh Institute of Applied Mathematics,\\
Russian Academy of Science, Miusskaya Pl. 4, Moscow, 125047, Russia}

\address[mysecondaryaddress]{School of Mathematics, Institute of Science, \\
Suranaree University of Technology, 30000, Thailand}

\date{\today}

\begin{abstract}

The two-dimensional shallow water equations in Eulerian and Lagrangain coordinates are considered.
Lagrangian and Hamiltonian formalism of the equations is given.
The transformations mapping the two-dimensional shallow water equations with
a circular or plane bottom into the gas dynamics equations of a
polytropic gas with polytropic exponent~$\gamma=2$ is represented.

Group properties of the equations are considered, and the group classification
for the case of the elliptic paraboloid bottom topography is performed.

The properties of the two-dimensional shallow water equations in Lagrangian coordinates
are discussed from the discretization point of view.
New invariant conservative finite-difference schemes for the equations and their one-dimensional reductions are constructed.
The schemes are derived either by extending the known one-dimensional schemes
or by direct algebraic construction based on some assumptions on the form of the energy conservation law.
Among the proposed schemes there are schemes possessing conservation laws of mass and energy.
\end{abstract}


\end{frontmatter}

\section{Introduction}


\bigskip


Shallow water equations are widely used to describe various physical
phenomena, for example, to study large-scale atmospheric and ocean
currents, to describe currents in the coastal zones of the seas and
oceans, to simulate tsunamis, the propagation of breakthrough waves
and tidal bores in rivers, the distribution of heavy gases and impurities
in the Earth's atmosphere.


One of the approaches of the analysis of nonlinear wave fluid motions
in rotating basins of various shapes is carried out in the framework
of the theory of shallow water \cite{bk:Vallis}. The rotating shallow
water model is a well-known nonlinear approximation used to describe
large-scale atmospheric and ocean currents. These equations make it
possible to provide important qualitative properties of the currents.
It should be mention here that there are many different approaches
for deriving shallow water equations.

\subsection{The shallow water equations in Eulerian and Lagrangian coordinates}

The hyperbolic shallow water equations were derived as the first-order
approximation with respect to the depth of the equations obtained
from the averaged inviscid incompressible fluid dynamics equations.
This approximation allows one to describe incompressible heavy fluid
flow with a free surface. These equations
have the following form
\begin{eqnarray}
u_t + u u_x + v u_y + h_x = H_x, \label{SWeulr1}
\\
v_t + u v_x + v v_y + h_y = H_y, \label{SWeulr2}
\\
h_t + (u h)_x + (v h)_y = 0, \label{SWeulr3}
\end{eqnarray}
where
$H$ is the height of the lower surface (the bottom topography),
$h = \eta + H$ is the total fluid thickness,
$\eta$ is the height of the upper free surface,
$(u,v)$ is the velocity,
$t$ is time,
$(x,y)$ are the Eulerian coordinates.

\begin{figure}[ht]
  \centering
  \includegraphics[width=0.25\linewidth]{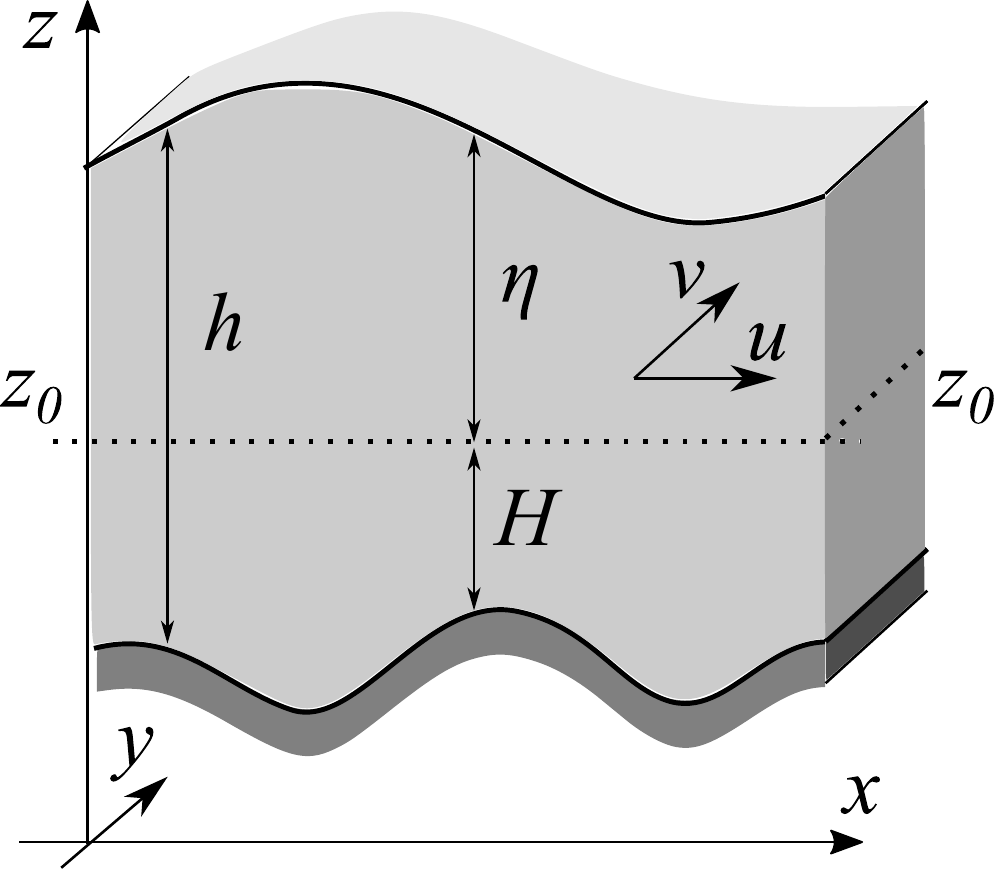}
  \caption{The notation in Eulerian coordinates}
  \label{fig:notation}
\end{figure}


Introducing Lagrangian coordinates $(t,\tilde{a},\tilde{b})$,
where the labels $\tilde{a}$ and $\tilde{b}$ denote initial coordinates of a particle,
the variables $x$ and $y$ become dependent
\[
x= \phi^1(t,\tilde{a},\tilde{b}),
\qquad
y = \phi^2(t,\tilde{a},\tilde{b}),
\]
and the relations between the Lagrangian~$(t, \tilde{a}, \tilde{b})$ and Eulerian~$(t,x,y)$ coordinates are
defined by the equations
\begin{equation} \label{LagrVels}
\phi^1_t(t,\tilde{a},\tilde{b}) = u(t, \phi^1(t,\tilde{a},\tilde{b}), \phi^2(t,\tilde{a},\tilde{b})),
\qquad
\phi^2_t(t,\tilde{a},\tilde{b}) = v(t, \phi^1(t,\tilde{a},\tilde{b}), \phi^2(t,\tilde{a},\tilde{b})).
\end{equation}
The conservation law of mass~(\ref{SWeulr3}) provides the relation~\cite{bk:Sedov[mss]}
\begin{equation} \label{massCL_Sedov}
{h} = {h}_0 / \tilde{J},
\end{equation}
where ${h}_0(\tilde{a},\tilde{b}) > 0$ is the function of integration,
and
\begin{equation}
\tilde{J}=\phi^1_{{\tilde{a}}}\phi^2_{{\tilde{b}}}-\phi^1_{{\tilde{b}}}\phi^2_{{\tilde{a}}} \neq 0.
\end{equation}
Applying the change
\begin{equation}
a = f^1(\tilde{a}, \tilde{b}),
\qquad
b = f^2(\tilde{a}, \tilde{b}),
\end{equation}
one finds that
\begin{equation}
\tilde{J}= (f^1_{\tilde{a}} f^2_{\tilde{b}} - f^1_{\tilde{b}} f^2_{\tilde{a}}) J.
\end{equation}
where
\begin{equation}\label{jac}
J=\phi^1_{{a}}\phi^2_{{b}}-\phi^1_{{b}}\phi^2_{{a}} \neq 0.
\end{equation}
Hence, choosing $f^1(\tilde{a}, \tilde{b})$ and $f^2(\tilde{a}, \tilde{b})$ such that
\begin{equation}
f^1_{\tilde{a}} f^2_{\tilde{b}} - f^1_{\tilde{b}} f^2_{\tilde{a}} = 2 {h}_0,
\end{equation}
one derives that
\begin{equation} \label{massCL_Sedov_Eqiv}
{h}(t,a,b) = H + \eta = 2 J^{-1}(t,a,b).
\end{equation}
Following the one-dimensional case, the coordinates $(t, a, b)$ are called the mass Lagrangian coordinates.
As there is no ambiguity, the sign $\tilde{\,}$ is further omitted.

\medskip

Finally, we rewrite~(\ref{SWeulr1}) and (\ref{SWeulr2})
in Lagrangian coordinates as
\begin{equation}
\begin{array}{c}
{{x}_{tt}}
+2J^{-3}(x_{{a}}y_{{a}}y_{{b}{b}}-(y_{{b}}x_{{a}}+x_{{b}}y_{{a}})y_{{a}{b}}%
+x_{{b}}y_{{a}{a}}y_{{b}}-x_{{a}{a}}y_{{b}}^{2}+2x_{{a}{b}}y_{{a}}y_{{b}}-x_{{b}{b}}y_{{a}}^{2})=H_{x},\\[2ex]
{{y}_{tt}}
+2J^{-3}(x_{{b}}y_{{b}}x_{{a}{a}}-(x_{{a}}y_{{b}}+x_{{b}}y_{{a}})x_{{a}{b}}%
+x_{{a}}y_{{a}}x_{{b}{b}}-x_{{b}}^{2}y_{{a}{a}}+2x_{{a}}x_{{b}}y_{{a}{b}}-x_{{a}}^{2}y_{{b}{b}})=H_{y}.
\end{array}\label{eq:raw}
\end{equation}

\subsection{Commutativity of Eulerian and Lagrangian derivatives}

The Lagrangian derivatives $D_a$, $D_b$ and $D_t^L$
can be defined through the Eulerian ones $D_t^E$, $D_x$ and $D_y$ as follows~\cite{bk:SamarskyPopov_book[1992]}
\begin{equation} \label{LagrDerivs}
D_t = D_t^E + u D_x + v D_y,
\qquad
D_a = \phi^1_a D_x + \phi^2_a D_y,
\qquad
D_b = \phi^1_b D_x + \phi^2_b D_y,
\end{equation}
where the total derivatives in Eulerian coordinates are defined as follows
\[
\def\arraystretch{1.75}
\begin{array}{l}
\displaystyle
D_t^E = \ppartial{t} + u_t \ppartial{u} + u_{tt} \ppartial{u_t} + u_{tx}\ppartial{u_x} + u_{ty}\ppartial{u_y} + \dots,
\\
\displaystyle
D_x = \ppartial{x} + u_x \ppartial{u} + u_{tx} \ppartial{u_t} + u_{xx}\ppartial{u_x} + u_{xy}\ppartial{u_y} + \dots,
\\
\displaystyle
D_y = \ppartial{y} + u_y \ppartial{u} + u_{ty} \ppartial{u_t} + u_{xy}\ppartial{u_x} + u_{yy}\ppartial{u_y} + \dots\,.
\end{array}
\]
From the latter relations it follows that the
Eulerian derivatives with respect to $x$ and $y$ are
\begin{equation} \label{EulerDerivsThroughLagr}
D_x = \frac{\phi^2_b D_a - \phi^2_a D_b}{J},
\qquad
D_y = \frac{\phi^1_a D_b - \phi^1_b D_a}{J},
\end{equation}
where $J$ is given by equation (\ref{jac}).

Here and further on for the sake of brevity we
write $\phi^1 \equiv x$ and $\phi^2 \equiv y$ keeping in mind that $x$ and $y$ are the coordinates of a Lagrangian particle.
Also we denote Lagrangian derivatives of a quantity $f$ as $f_t$, $f_a$ and $f_b$,
and its Eulerian derivatives as $f^E_t$, $f_x$ and $f_y$.

\medskip

Notice that the Lagrangian derivative $D_t^L$ does not
commute with the Eulerian derivatives $D_x$ and $D_y$:
\begin{equation}
\def\arraystretch{1.75}
\begin{array}{c}
[D_t^L, D_x] = D_x D_t^L - D_t^L D_x = u_x D_x + v_x D_y \neq 0,
\\ \null
[D_t^L, D_y] = D_y D_t^L - D_t^L D_y = u_y D_x + v_y D_y \neq 0.
\end{array}
\end{equation}
We now show that the Lagrangian derivatives~(\ref{LagrDerivs})
do commute in any order.

First,
\begin{multline}
[D_a, D_b]
= [(x_b)_y y_a + (x_b)_x x_a - (x_a)_x x_b - (x_a)_y y_b] D_x
\\
- [(y_a)_x x_b + (y_a)_y y_b - (y_b)_x x_a - (y_b)_y y_a] D_y.
\end{multline}
By means of~(\ref{EulerDerivsThroughLagr}),
\begin{multline}
(x_b)_y y_a + (x_b)_x x_a - (x_a)_x x_b - (x_a)_y y_b
 = \frac{1}{{J}}[
    (x_a x_{bb} - x_b x_{ab}) y_a
    + (x_{ab} y_b - x_{bb} y_a) x_a
    \\
    - (x_{aa} y_b - x_{ab} y_a) x_b
    - (x_a x_{ab} - x_b x_{aa}) y_b
 ] = 0.
\end{multline}
The similar way one shows that
\[
(y_a)_x x_b + (y_a)_y y_b - (y_b)_x x_a - (y_b)_y y_a = 0.
\]
Thus, the operators $D_a$ and $D_b$ commute on smooth enough solutions of the considered equations.

Next,
\begin{multline}
[D_t^L, D_a]
= (x_a u_x + y_a u_y - u (x_a)_x - v (x_a)_y - (x_a)_t^E) D_x
+ (x_a v_x + y_a v_y - u (y_a)_x - v (y_a)_y - (y_a)_t^E) D_y
\\
= (x_a (x_t)_x + y_a (x_t)_y - x_{at}) D_x
+ (x_a (y_t)_x + y_a (y_t)_y - y_{at}) D_y
\\
= \left[\frac{1}{{J}}\left(
    x_a (x_{ta} y_b - x_{tb} y_a)
    + y_a (x_a x_{tb} -x_b x_{ta})
\right) - x_{at} \right] D_x
\\
+ \left[\frac{1}{{J}}\left(
    x_a (y_{ta} y_b - y_{tb} y_a)
    + y_a (x_a y_{tb} -x_b y_{ta})
\right) - y_{at} \right] D_y
\\
= \left[\frac{1}{{J}}
    x_{ta} (x_a y_b - x_b y_a) - x_{at}
\right] D_x
+ \left[\frac{1}{{J}}
    y_{ta} (x_a y_b - x_b y_a) - y_{at}
\right] D_y
\\
=
(x_{ta}  - x_{at}) D_x
+ (y_{ta}  - y_{at}) D_y
= 0.
\end{multline}
The similar way one shows that $[D_t, D_b] = 0$.

Thus, it was shown that the operators $D_t^L$, $D_a$ and $D_b$ commute on smooth enough solutions of the system.

\subsection{The group analysis of the shallow water equations
(\ref{SWeulr1})--(\ref{SWeulr3})}

Group analysis of the one-dimensional shallow water equations has
been applied in numerous papers. A comprehensive review of these results
can be found in \cite{bk:KaptsovMeleshko2019_2}\footnote{See also literature therein.}.
Among these, we mention the papers \cite{bk:AksenovDruzhkov[2016],bk:KaptsovMeleshko2019_2},
where variable bottom topography was considered. In \cite{bk:AksenovDruzhkov[2016]}
the study was performed in Eulerian coordinates. The conservation
laws were derived by the direct method, which had been applied earlier
to the gas dynamics equations \cite{bk:TerentievShmyglevski,bk:Shmyglevski}.
On the other hand, in \cite{bk:KaptsovMeleshko2019_2} the shallow
water equations were studied in Lagrangian coordinates.

The group analysis method has been applied to the two-dimensional
shallow water equations in \cite{bk:LeviNucciRogersWinternitz,bk:BilaMansfieldClarkson2006,bk:Chesnokov2008,bk:Chesnokov2009,bk:Chesnokov2011,bk:AksenovDruzhkov[2018_1],bk:AksenovDruzhkov[2018_2],bk:AksenovDruzhkov[2019],bk:MeleshkoSamatova2019_Albena}.

In \cite{bk:AksenovDruzhkov[2018_1],bk:AksenovDruzhkov[2018_2],bk:AksenovDruzhkov[2019]}
group classification and conservation laws of the two-dimensional
shallow-water equations over an uneven bottom in the absence of a
Coriolis force ($f=0$) were studied. For finding conservation laws
the authors of \cite{bk:AksenovDruzhkov[2018_1],bk:AksenovDruzhkov[2018_2],bk:AksenovDruzhkov[2019]}
used the same approach as in \cite{bk:AksenovDruzhkov[2016]}.

In the papers \cite{bk:LeviNucciRogersWinternitz,bk:BilaMansfieldClarkson2006,bk:Chesnokov2008,bk:Chesnokov2009,bk:Chesnokov2011}
the Coriolis parameter was assumed to be constant, $f=\textrm{const}$, whereas
in \cite{bk:MeleshkoSamatova2019_Albena} $f=f_{0}+\beta y$ ($\beta\neq0$).
The authors of \cite{bk:LeviNucciRogersWinternitz} studied group
properties of the shallow water equations with the elliptic paraboloid
\[
H=Ax^{2}+By^{2},\ \ (A>0,\ B>0),
\]
and it was shown that if $A\neq B$, then the admitted Lie algebra
is six-dimensional, while if the bottom is a circular paraboloid,
then it is nine-dimensional. It was noted in \cite{bk:Chesnokov2011}
that for a circular bottom the admitted Lie algebra is isomorphic
to the Lie algebra admitted by the classical shallow water equations
with flat bottom $H=\textrm{const}$. This allowed to guess and then to
prove that there is a change of the dependent and independent variables
such that these two systems of equations are locally equivalent. For
the particular ($f\neq0$) case $A=B=0$, this property had been proven
by the same author earlier in \cite{bk:Chesnokov2008,bk:Chesnokov2009}.
In \cite{bk:LeviNucciRogersWinternitz,bk:Chesnokov2009,bk:Chesnokov2008,bk:Chesnokov2011},
the study was performed in Eulerian coordinates, whereas in \cite{bk:BilaMansfieldClarkson2006,bk:MeleshkoSamatova2019_Albena},
the two-dimensional shallow water equations were considered in Lagrangian
coordinates. In \cite{bk:BilaMansfieldClarkson2006}, group properties
of the two-dimensional shallow water equations over a flat bottom
($H=\textrm{const}$) were studied. Using a Lagrangian of the form presented
in \cite{bk:Salmon1983}, the authors of \cite{bk:BilaMansfieldClarkson2006}
constructed conservation laws by applying Noether's theorem. According
to \cite{bk:Chesnokov2008,bk:Chesnokov2009} the shallow water equations ($H=\textrm{const}$)
analysed in \cite{bk:BilaMansfieldClarkson2006} are equivalent to
the gas dynamics equations of an isentropic flow of a polytropic gas
for the exponent $\gamma=2$. The group properties and conservation
laws of the two-dimensional gas dynamics equations in Lagrangian coordinates
were studied in \cite{bk:KaptsovMeleshko2019}.

One advantage of choosing Lagragian coordinates for the study of the
shallow water equations is that equations (\ref{SWeulr1})--(\ref{SWeulr3}) have
a variational structure: choosing the Lagrangian\footnote{The authors of \cite{bk:BilaMansfieldClarkson2006} used a different
Lagrangian \cite{bk:Salmon1983}.}
\begin{equation}
{\cal L}=\frac{1}{2}({x}_t^{2}+{y}_t^{2})-(J^{-1}-H),\label{eq:Lagrangian}
\end{equation}
the shallow water equations (\ref{SWeulr1})--(\ref{SWeulr3}) turn to be the Euler-Lagrange
equations ${\displaystyle \frac{\delta{\cal L}}{\delta x}=0}$ and
${\displaystyle \frac{\delta{\cal L}}{\delta y}=0}$, where ${\displaystyle \frac{\delta}{\delta x}}$
and ${\displaystyle \frac{\delta}{\delta y}}$ are variational derivatives.
This variational structure allowed the authors of \cite{bk:BilaMansfieldClarkson2006,bk:KaptsovMeleshko2019}
to apply Noether's theorem \cite{bk:Noether[1918]} for deriving
conservation laws.

Application of the Hamiltonian principle to fluid dynamics in Eulerian
coordinates can be found in \cite{bk:Salmon1988,bk:Salmon[1998]}.

The group classification of the shallow water equations with constant
Coriolis parameter and a variable bottom topography was studied in
\cite{bk:Meleshko2020}.

The group classification of the shallow water equations without Coriolis
force in Eulerian coordinates and variable bottom topography was studied
in \cite{bk:BihloPoltavetsPopovych2020}.

\bigskip

This paper is organized as follows. Lagrangian and Hamiltonian
formalism of equations (\ref{SWeulr1})--(\ref{SWeulr3}) is given in the next section.
Preliminary analysis of equations (\ref{SWeulr1})--(\ref{SWeulr3}) is presented
in Section~\ref{sec:Prelim}, where the transformations mapping the two-dimensional
shallow water equations with 
a circular or plane bottom into the gas dynamics equations of a
polytropic gas with polytropic exponent $\gamma=2$ (equations (\ref{SWeulr1})--(\ref{SWeulr3}))
with $H=\textrm{const}$) are found. Group properties of equations (\ref{SWeulr1})--(\ref{SWeulr3})
are described in Section~\ref{sec:GroupClassif}.

The shallow water equations in Lagrangian coordinates~(\ref{eq:raw})
are discussed from the discretization point of view in Section~\ref{sec:PrelimDiscrete}.
Their symmetry properties are considered in more detail and some conservation laws of the equations are provided.
It is also shown that equations~(\ref{eq:raw}) can be rewritten in conservative form.
In Section~\ref{sec:Discr}, invariant finite-difference schemes for the two-dimensional shallow water equations in Lagrangian coordinates and their reductions are constructed. Two different approaches for construction
of such schemes are proposed: 1)~construction by extending the known one-dimensional schemes,
and 2)~direct algebraic construction assuming the general form of the conservation law of energy is known.
The results are summarized in Conclusion.

\section{Lagrangian and Hamiltonian formalism of equations (\ref{SWeulr1})--(\ref{SWeulr3}) }
\label{sec:LagrHamFormalism}

The Lagrangian (\ref{eq:Lagrangian}) has the form
\begin{equation}
{\cal L}({x}_t,{y}_t,x,y,x_{{a}},x_{{b}},y_{{a}},y_{{b}})=\frac{1}{2}({x}_t^{2}+{y}_t^{2})+g(x,y,x_{{a}},x_{{b}},y_{{a}},y_{{b}}).\label{eq:Lagrangian_gen}
\end{equation}
The Euler-Lagrange equations are
\begin{equation}
{x}_{tt}=\frac{\delta g}{\delta x},\,\,\,{y}_{tt}=\frac{\delta g}{\delta y}.\label{eq:1}
\end{equation}
Introducing the variables
\[
c_{1}={x}_t,\,\,\,c_{2}=y_t,
\]
the Lagrangian ${\cal L}$ becomes
\[
{\cal L}(c_{1},c_{2},x,y,x_{{a}},x_{{b}},y_{{a}},y_{{b}})=\frac{1}{2}(c_{1}^{2}+c_{2}^{2})+g(x,y,x_{{a}},x_{{b}},y_{{a}},y_{{b}}).
\]
The Euler-Lagrange equations (\ref{eq:2}) in the Lagrangian formalism
can be rewritten in the form
\begin{equation}
({a}_{1})_t=\frac{\delta{\cal L}}{\delta x},\,\,\,(a_{2})_t=\frac{\delta{\cal L}}{\delta y},\,\,\,x_t=c_{1},\,\,\,y_t=c_{2},\label{eq:2}
\end{equation}
where $c_{1}$ and $c_{2}$ are found from the equations
\[
a_{1}=\frac{\delta{\cal L}}{\delta c_{1}}=c_{1},\,\,\,a_{2}=\frac{\delta{\cal L}}{\delta c_{2}}=c_{2}.
\]

As the Lagrangian (\ref{eq:Lagrangian_gen}) is nonsingular \cite{bk:Mokhov},
then one can derive the Hamiltonian form as follows \cite{bk:Ostrogradsky}.

Using the Legendre transformation
\[
H={x}_t{\cal L}_{{x}_t}+{y_t}{\cal L}_{{y_t}}-{\cal L}=\frac{1}{2}({x}_t^{2}+{y}_t^{2})-g,
\]
the Hamiltonian becomes
\begin{equation}
\begin{array}{c}
H(a_{1},a_{2},x,y,x_{{a}},x_{{b}},y_{{a}},y_{{b}})=\frac{1}{2}(a_{1}^{2}+a_{2}^{2})-g(x,y,x_{{a}},x_{{b}},y_{{a}},y_{{b}}).\end{array}\label{eq:3}
\end{equation}
The Hamiltonian equations are
\begin{equation}
{x_t}=\frac{\delta H}{\delta a_{1}},\,\,\,{y_t}=\frac{\delta H}{\delta a_{2}},\,\,\,({a}_{1})_t=-\frac{\delta H}{\delta x},\,\,\,({a}_{2})_t=-\frac{\delta H}{\delta y}.\label{eq:Hamiltonian}
\end{equation}
Substituting the Hamiltonian (\ref{eq:3}) into (\ref{eq:Hamiltonian}),
the Hamiltonian equations become
\[
{x_t}=a_{1},\,\,\,{y_t}=a_{2},\,\,\,({a}_{1})_t=\frac{\delta g}{\delta x},\,\,\,({a}_{2})_t=\frac{\delta g}{\delta y}.
\]
Hence,
\[
a_{1}={x_t},\ \ a_{2}={y_t},
\]
and one notes that equations (\ref{eq:Hamiltonian}) coincide with
(\ref{eq:1}).

\section{Preliminary consideration}
\label{sec:Prelim}

As the gas dynamics equations have been extensively studied, before
proceeding to the group classification, we show that for particular
bottoms the shallow water equations (\ref{SWeulr1})--(\ref{SWeulr3}) can be reduced
to the gas dynamics equations ($f=0,\ H=\textrm{const}$) of a polytropic gas
with the exponent $\gamma=2$. In \citep{bk:Meleshko2020}\footnote{In \citep{bk:Meleshko2020}, the author considered the shallow water
equations with a constant Coriolis parameter $f\neq0$. However, one
can check that the found there transformations are also valid and
for equations with $f=0$. } it was found transformations mapping the shallow water equations
(\ref{SWeulr1})--(\ref{SWeulr3}) with $H=p(x^{2}+y^{2})$ and $H=q_{1}x+q_{2}y$
into equations (\ref{SWeulr1})--(\ref{SWeulr3}) with horizontal bottom $H=0$.

\begin{rem}

Particular case of transformations found in \citep{bk:Meleshko2020}
is the change
\[
f^{t}=t^{-1},\,\,\,f^{x}=2t^{-1}(x+y),\,\,\,f^{y}=2t^{-1}(x-y),
\]
\[
f^{h}=8ht^{2},\,\,\,f^{u}=-2t(u+v)+2(x+y),\,\,\,f^{v}=2t(-u+v)+2(x-y),
\]
which leaves equations (\ref{SWeulr1})--(\ref{SWeulr3}) invariant.

\end{rem} 

\section{Group classification}
\label{sec:GroupClassif}

There is vast literature dedicated to the group classification of
classes of differential equations. A comprehensive review can be found,
for example, in \cite{bk:OpanasenkoBihloPopovych2017,bk:OpanasenkoBoykoPopovych2018}\footnote{See also literature therein.}.
In the present paper we use the classical approaches \cite{bk:Ovsiannikov1978}.

Equations (\ref{SWeulr1})--(\ref{SWeulr3}) contain the arbitrary function $H(x,y)$.
The first step in group classification is to find transformations
that change the arbitrary elements while preserving the differential
structure of the equations themselves. Such transformations are called
equivalence transformations. The group classification is considered
with respect to equivalence transformations.

\subsection{Equivalence group}

A generator of an equivalence Lie group \cite{bk:Ovsiannikov1978}
is assumed to be in the form \cite{bk:Meleshko[2005]}
\[
X^{e}=\zeta\partial_{t}+\zeta^{{a}}\partial_{{a}}+\zeta^{{b}}\partial_{{b}}+\zeta^{x}\partial_{x}+\zeta^{y}\partial_{y}+\zeta^{f_{0}}\partial_{f_{0}}+\zeta^{\beta}\partial_{\beta}+\zeta^{H}\partial_{H}
\]
where all coefficients of the generator depend on $(t,{a},{b},x,y,f,H)$.
Applying the prolonged generator to the system consisting of equation
(\ref{SWeulr1})--(\ref{SWeulr3}) and the equations
\[
H_{t}=0,\,\,\,H_{{a}}=0,\,\,\,H_{{b}}=0,
\]
and splitting them with respect to the parametric derivatives, one
obtains an overdetermined system of partial differential equations.
Solving this system, one finds the equivalence group. The equivalence
group corresponds to the generators:
\[
X_{1}^{e}=\partial_{x},\,\,\,X_{2}^{e}=\partial_{y},\,\,\,X_{3}^{e}=y\partial_{x}-x\partial_{y},\,\,\,X_{\psi}^{e}=\psi_{{b}}\partial_{{a}}-\psi_{{a}}\partial_{{b}},
\]
\[
X_{4}^{e}=\partial_{t},\,\,\,X_{5}^{e}=2{a}\partial_{{a}}+t\partial_{t}+x\partial_{x}+y\partial_{y},
\]
\[
X_{6}^{e}=\partial_{H},\,\,\,X_{7}^{e}=2t\partial_{t}+x\partial_{x}+y\partial_{y}-2H\partial_{H}.
\]

There are also two involutions
\[
\begin{array}{c}
E_{1}:\ \ \ \ \ t\to-t;\\
E_{2}:\ \ \ \ \ x\to-x,\ \ y\to-y;
\end{array}
\]

\subsection{Classification of Admitted Lie groups}

An admitted generator is sought in the form
\[
X=\zeta\partial_{t}+\zeta^{{a}}\partial_{{a}}+\zeta^{{b}}\partial_{{b}}+\zeta^{x}\partial_{x}+\zeta^{y}\partial_{y},
\]
where all coefficients depend on $(t,{a},{b},x,y)$.

Applying the prolonged generator to equations (\ref{SWeulr1})--(\ref{SWeulr3}),
the determining equations are reduced to the study of the classifying
equation
\begin{equation}
(\alpha x+\beta y+\gamma_{1})H_{x}+(-\beta x+\alpha y+\gamma_{2})H_{y}=2\gamma H+q(x^{2}+y^{2})+q_{1}x+q_{2}y+q_{0}.\label{classif}
\end{equation}
where
\[
\alpha=\zeta^{\prime}+2k_{1},\ \ \beta=2k_{4},\ \ \gamma=2(\zeta^{\prime}-2k_{1}),\ \ q=-\frac{1}{2}\zeta^{\prime\prime\prime},
\]
\[
\gamma_{1}=2\zeta_{1},\ \ \gamma_{2}=2\zeta_{2},\ \ q_{1}=-2\zeta_{1}^{\prime\prime},\ \ q_{2}=-2\zeta_{2}^{\prime\prime},\ \ q_{0}=g,
\]
\[
\zeta^{{a}}=-\psi_{{b}}+4k_{1}{a},\,\,\,\zeta^{{b}}=\psi_{{a}},\,\,\,
\]
\[
\zeta^{x}=\frac{1}{2}\zeta^{\prime}x+k_{1}x+k_{4}y+\zeta_{1},\,\,\,\zeta^{y}=\frac{1}{2}\zeta^{\prime}y+k_{1}y-k_{4}x+\zeta_{2},
\]
and the functions $\zeta(t)$, $\zeta_{1}(t)$, $\zeta_{2}(t)$, $g(t)$
and $\psi=\psi({a},{b})$ are arbitrary functions of their arguments.

The kernel of admitted Lie algebras, which is admitted for all cases
of the function $H(x,y)$, consists of the generators
\begin{equation}
X_{1}=\partial_{t},\qquad X_{2}=-\psi_{{b}}\partial_{{a}}+\psi_{{a}}\partial_{{b}}.\label{kernel}
\end{equation}
Extensions of the kernel occur for specific functions $H(x,y)$. Consideration
of these cases leads to the analysis similar to applied in \citep{bk:Meleshko2020}\footnote{In Eulerian coordinates group classification of equations (\ref{SWeulr1})--(\ref{SWeulr3})
was done in \citep{bk:BihloPoltavetsPopovych2020}.}.

\smallskip

We restrict ourselves in this paper with the case~$H=p x^{2}+2cxy+by^{2}+q_{1}x+q_{2}y+q_{0}$.

Using an orthogonal transformation one can reduce the function $H(x,y)$
to the canonical form:
\begin{equation}\label{btmSqr}
H(x,y)=\lambda_{1}x^{2}+\lambda_{2}y^{2}+q_{1}x+q_{2}y+q_{0}.
\end{equation}

Notice that for $\lambda_{1}\neq0$, by virtue of the equivalence
transformations corresponding to the shifts of $x$ and $H$, one
can assume that $q_{1}=0$. Similar, one can assume that $q_{2}=0$
if $\lambda_{2}\neq0$. Using the shift $H$, one can assume that
$q_{0}=0$. By virtue of the preliminary study, this allows us to
state the following theorem.

\begin{thm}
\label{th:th1}
If the bottom has the form
\begin{equation}
H(x,y)=p (x^{2}+y^{2})+q_{1}x+q_{2}y+q_{0},\label{eq:Dec03.1}
\end{equation}
where $p$ and $q_{i}$ ($i=0,1,2$) are constant, then system of
the shallow water equations (\ref{SWeulr1})--(\ref{SWeulr3}) can be reduced to
the gas dynamics equations\footnote{Equations (\ref{SWeulr1})--(\ref{SWeulr3}) with $f=0$ and $H=\textrm{const}$.}
of a polytropic gas with the exponent $\gamma=2$. \end{thm}

Thus, the group classification of equations (\ref{SWeulr1})--(\ref{SWeulr3}) with
the bottom (\ref{btmSqr}) is restricted to the following cases
\begin{enumerate}[a)]
  \item $\lambda_{1}\neq\lambda_{2}$;
  \item $\lambda_{1} = \lambda_{2} = q_1 = q_2 = q_0 = 0$.
\end{enumerate}

Calculations in the first case are straight forward and the classification
results are listed in Table~\ref{tab:classifQuadric}. In the first
column of the table different forms of the function~$H$ are given.
The corresponding extensions of the kernel~(\ref{kernel}) and constraints
on the constants and arbitrary functions are presented in the second
and the third columns of the table.

\begin{center}
\def\arraystretch{1.25}
\setlength{\LTleft}{0pt} \setlength{\LTright}{0pt} %
\begin{longtable}{@{\extracolsep{\fill}}C{0.5cm}C{2.5cm}C{7.5cm}C{4cm}}
\caption{Classification results $H(x,y)$}
\label{tab:classifQuadric}\\
\hline
\#  & {$H$}  & {Extension}  & {Conditions} \\
\hline
\endfirsthead
\hline
 &  & {\textit{Table \thetable\ (continued)}}  & \\
\hline
\#  & {$H$}  & {Extension}  & {Conditions}\\
\hline
\endhead
\hline
\hline
\rowno  & \classifcase{dop74a1} $\lambda_{1}x^{2}+\lambda_{2}y^{2}$  & $\begin{array}{c}
x\partial_{x}+y\partial_{y}+4a\partial_{a},\\
\zeta_{1}(t)\partial_{x}+\zeta_{2}(t)\partial_{y}
\end{array}$
  & $\begin{array}{c}
\lambda_{1}\lambda_{2}(\lambda_{1}-\lambda_{2})\neq0,\\
\zeta_{1}^{\prime\prime}=2\lambda_{1}\zeta_{1},\\
\zeta_{2}^{\prime\prime}=2\lambda_{2}\zeta_{2}
\end{array}$ 
\\
\hline
\rowno  & \classifcase{dop74a2} $\lambda_{1}x^{2}+q_{2}y$
  & $\begin{array}{c}
4x\partial_{x}+16 a\partial_{a}+(2y-q_{2}t^{2})\partial_{y},\\
t\partial_{y},\ \ \partial_{y},\,\,\,\zeta(t)\partial_{x},
\end{array}$
  &$\begin{array}{c}
  \lambda_{1}\neq 0,\\
   \zeta_{1}^{\prime\prime}=2\lambda_{1}\zeta_{1}\end{array}$
\\
\hline
\end{longtable}
\par\end{center}

\global\long\def\arraystretch{1.25}

\begin{center}
\setlength{\LTleft}{0pt} \setlength{\LTright}{0pt}
\par\end{center}

\section{Preliminary analysis of the shallow water equations for constructing finite-difference schemes}
\label{sec:PrelimDiscrete}

\subsection{The general case $H=H(x,y)$}

For the further discretization of equations~(\ref{eq:raw}) it is more useful to consider
them in the following form
\begin{equation} \label{SWdiv1}
F_1 = D_t({x}_t) + D_a\left( y_b J^{-2} \right) - D_b\left( y_a J^{-2} \right) - H_x = 0,
\end{equation}
\begin{equation} \label{SWdiv2}
F_2 = D_t({y}_t) - D_a\left( x_b J^{-2} \right) + D_b\left( x_a J^{-2} \right) - H_y = 0.
\end{equation}
Recall that $x$ and $y$ denote the coordinates of a Lagrangian particle.

\medskip


As it was shown above, the kernel of admitted Lie algebras, which is admitted by equations~(\ref{SWdiv1}) and~(\ref{SWdiv2})
for all cases of the function $H(x,y)$, consists of the generators
\begin{equation} \label{kernel}
X_{1}=\ppartial{t},
\qquad
X_{2}= \psi_{b}\ppartial{a} - \psi_{a}\ppartial{b},
\end{equation}
where $\psi(a, b)$ is an arbitrary function corresponding to relabelling of Lagrangian variables.

Notice that there are shifts
$\ppartial{a}$ and
$\ppartial{b}$,
inhomogeneous scaling
$a\ppartial{a} - b \ppartial{b}$
and rotation
$b \ppartial{a} - a \ppartial{b}$
among the particular forms of the generator~$X_{2}$.

\medskip

Equations~(\ref{SWdiv1}),~(\ref{SWdiv2}) possess the local conservation laws of mass, energy and momentum.

By means of equations~(\ref{LagrVels}) and~(\ref{massCL_Sedov_Eqiv})
one can write the Eulerian conservation law of mass~(\ref{SWeulr3})
in Lagrangian coordinates as
\begin{equation} \label{CLmass}
D_t(J) = D_t(x_{{a}}y_{{b}}-x_{{b}}y_{{a}}) = D_a({x}_t y_b - {y}_t x_b ) + D_b({y}_t x_a  - {x}_t y_a).
\end{equation}

The conservation law of energy which corresponds to the generator~$X_1=\ppartial{t}$
can be obtained with the help of Noether's theorem~\cite{bk:Noether[1918],bk:Ibragimov[1983]}
\begin{multline} \label{CLenergy}
    {x}_t \left[
        D_t({{x}_t}) + D_a\left( y_b J^{-2} \right) - D_b\left( y_a J^{-2} \right) - H_x
    \right]
    \\
    + {{y}_t} \left[
        D_t({{y}_t}) - D_a\left( x_b J^{-2} \right) + D_b\left( x_a J^{-2} \right) - H_y
    \right]
    \\
    = D_t\left[ \frac{1}{2}({{x}_t}^2 + {{y}_t}^2) + J^{-1} - H \right]
    + D_a\left[
        ({{x}_t}y_b - {{y}_t}x_b) J^{-2}
    \right]
    + D_b\left[
        ({{y}_t}x_a - {{x}_t}y_a) J^{-2}
    \right] = 0.
\end{multline}

The conservation law of momentum which corresponds
to the generators~$\ppartial{a}$ and~$\ppartial{b}$ is
\begin{multline} \label{CLmomentum}
    ({x}_a + x_b) \left[
        D_t({{x}_t}) + D_a\left( y_b J^{-2} \right) - D_b\left( y_a J^{-2} \right) - H_x
    \right]
    \\
    + ({y}_a + y_b) \left[
        D_t({{y}_t}) - D_a\left( x_b J^{-2} \right) + D_b\left( x_a J^{-2} \right) - H_y
    \right]
    \\
    = D_t\left[
        (x_a + x_b) x_t
        + (y_a + y_b) y_t
    \right]
    + D_a\left[
        2 J^{-1} - \frac{x_t^2 + y_t^2}{2} - H
    \right]
    + D_b\left[
        2 J^{-1} - \frac{x_t^2 + y_t^2}{2} - H
    \right] = 0.
\end{multline}


\begin{rem}
Equations (\ref{SWdiv1})--(\ref{SWdiv2}) can be reduced to the one-dimensional shallow water equation
\begin{equation} \label{scheme1d}
{{x}_{tt}} - \frac{2 x_{aa}}{x_a^3} - \tilde{H}^\prime(x) = 0,
\end{equation}
by means of the relations $x(t,a,b) = x(t,a)$, $y(t, a, b) \equiv b$.

The identity~(\ref{CLmass}) becomes
\begin{equation}
D_t(x_a) - D_a(x_t) = 0,
\end{equation}
and the Jacobian $J$ is just reduced to $x_a$ in this case.
\end{rem}

\subsection{The case of a horizontal bottom $H = \textrm{const}$}

For the further discretization purposes, here we consider the case of a horizontal bottom topography ($H=\textrm{const}$) in more detail.

In case $H=\textrm{const}$, the shallow water equations in Lagrangian coordinates are
\begin{equation} \label{SWdiv1Flat}
F_1^0 = {{x}_{tt}} + D_a\left( y_b J^{-2} \right) - D_b\left( y_a J^{-2} \right) = 0,
\end{equation}
\begin{equation} \label{SWdiv2Flat}
F_2^0 = {{y}_{tt}} - D_a\left( x_b J^{-2} \right) + D_b\left( x_a J^{-2} \right) = 0,
\end{equation}
and the admitted Lie algebra is the same as for the
two-dimensional gas isentropic flows for the polytropic constant~$\gamma = 2$
~\cite{bk:Ibragimov[1983],bk:HandbookLie_v2}, i.e.,
the extension of the kernel~(\ref{kernel}) is
\begin{equation} \label{LieAlgFlat}
\def\arraystretch{2}
\begin{array}{c}
    \displaystyle
    Y_1 = \ppartial{x},
    \qquad
    Y_2 = \ppartial{y},
    \quad
    Y_3 = t \ppartial{x},
    \qquad
    Y_4 = t \ppartial{y},
    \\
    \displaystyle
    Y_5 = y \ppartial{x} - x \ppartial{y},
    \qquad
    Y_6 = t^2 \ppartial{t} + t x \ppartial{x} + t y \ppartial{y},
    \\
    \displaystyle
    Y_7 = 2 t \ppartial{t} + x \ppartial{x} + y \ppartial{y},
    \qquad
    Y_8 = 2 a \ppartial{a} + 2 b \ppartial{b} + x \ppartial{x} + y \ppartial{y}.
\end{array}
\end{equation}

Equations~(\ref{SWdiv1Flat}),~(\ref{SWdiv2Flat}) are ``weak-invariant''
with respect to the generators $Y_5$--$Y_8$, i.e.,
they satisfy the infinitesimal criterion on solutions only:
\[
\displaystyle
Y_k (F^0_j) |_{F^0_1 = 0, \,\, F^0_2 = 0}= 0,
\qquad
j = 1, 2,
\quad
k = 5, 6,\dots , 8.
\]

\begin{rem} \label{rem:jac}
Jacobian~(\ref{jac}) is a differential invariant of the generators $X_1$, $X_2$, $Y_1$--$Y_5$.
\end{rem}

The conservation laws for the case $H = \textrm{const}$ are listed in~\cite{bk:KaptsovMeleshko2019_2}.
Among them there is the center-of-mass law
\begin{equation} \label{CLcom}
D_t(t ({{x}_t} +{{y}_t}) - x - y) + D_a\left( t (y_b - x_b) J^{-2} \right) + D_b\left( t (x_a- y_a) J^{-2} \right) = 0.
\end{equation}

\section{Discretization of the shallow water equations in Lagrangian coordinates}
\label{sec:Discr}

It is often more suitable to construct invariant finite-difference schemes for equations of continuum mechanics in Lagrangian coordinates then in Eulerian ones~\cite{bk:Dorodnitsyn[2011]}, as the generators admitted by the equations in Lagrangian coordinates typically preserve uniformness and orthogonality of the corresponding finite difference meshes. Such invariant conservative schemes have been successfully constructed by the authors in~\cite{bk:DorKapSW2020,bk:DorKapSW_JMP_2020,bk:DorKapMelGN2020}.

\subsection{Notation}

\begin{figure}
  \centering
  \includegraphics[width=0.2\linewidth]{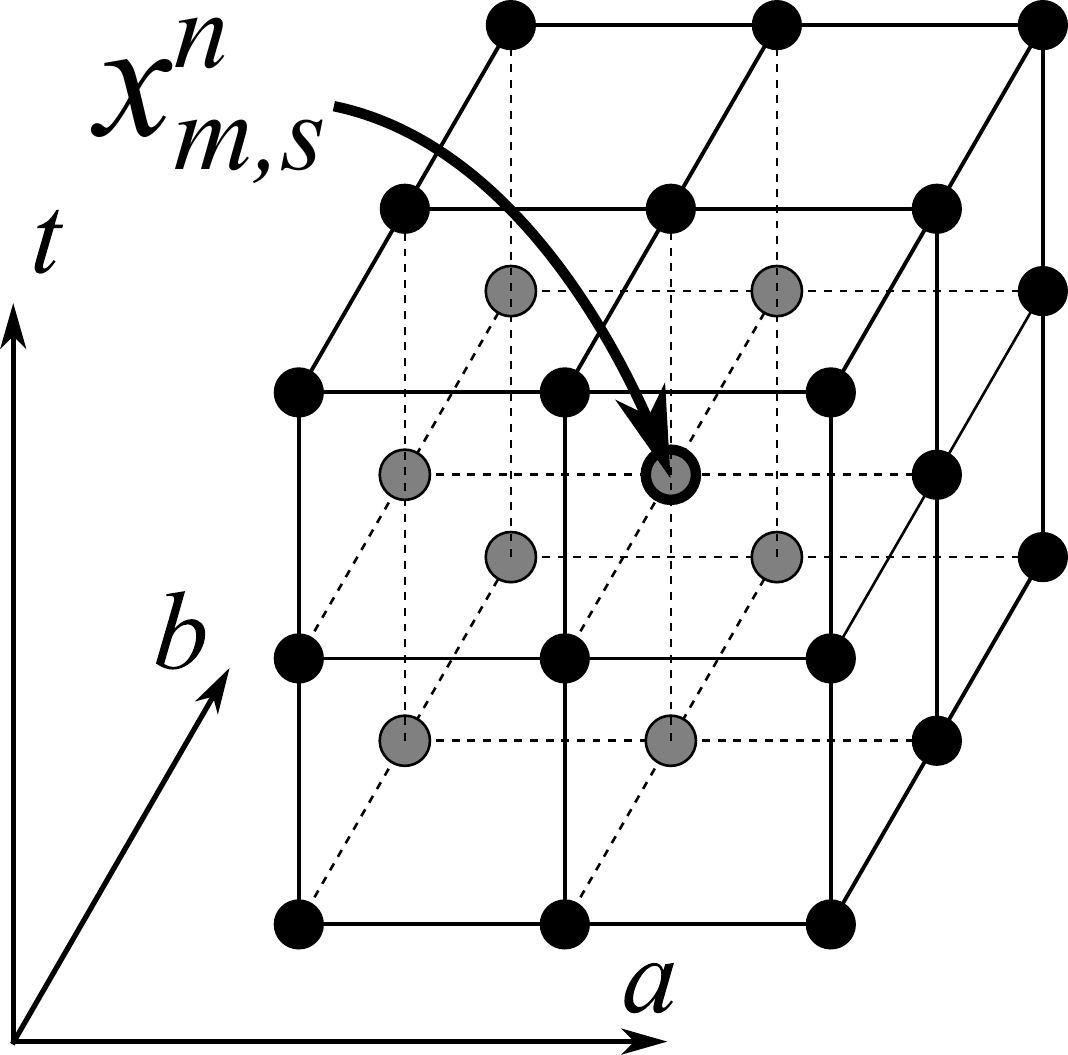}
  \caption{The 27-point stencil, Lagrangian coordinates}
  \label{fig:27pt}
\end{figure}

Following this approach, we consider discretizations in Lagrangian coordinates in $3 + 3 + 3 + 27 + 27 = 63$ variables on 27-point stencil~(see Figure~\ref{fig:27pt}), i.e.,
\begin{equation}
t_{n+i} \equiv t^{n+i}_{m+j, s+k},
\quad
a_{m+j} \equiv a^{n+i}_{m+j, s+k},
\quad
b_{s+k} \equiv b^{n+i}_{m+j, s+k},
\end{equation}
\[
x^{n+i}_{m+j, s+k},
\quad
y^{n+i}_{m+j, s+k},
\qquad
i, j, k = -1, 0, 1,
\]
or, in alternative notation,
\begin{equation}
\def\arraystretch{1.5}
\begin{array}{c}
    \displaystyle
    \check{t}, t, \hat{t},
    \quad
    a_-, a, a_+,
    \quad
    {}^-b, b, {}^+b,
    \\
    \displaystyle
    {}^-\check{x}_-,
    \check{x}_-,
    \check{x},
    \check{x}_+,
    {}^+\check{x}
    ,
    \dots,
    {}^+\hat{x}_+,
    \\
    \displaystyle
    {}^-\check{y}_-,
    \check{y}_-,
    \check{y},
    \check{y}_+,
    {}^+\check{y}
    ,
    \dots,
    {}^+\hat{y}_+,
\end{array}
\end{equation}
where
\[
    \hat{z} = z^{n + 1}_{m,s},
    \quad
    \check{z} = z^{n - 1}_{m,s},
    \quad
    z_{\pm} = z^n_{m \pm 1,s},
    \quad
    {}^{\pm}z = z^n_{m,s\pm 1}
    \quad
    \textrm{for}
    \quad
    z \in \{ t, a, b, x ,y \}.
\]

\smallskip

The standard finite-difference shift operators are defined as follows
\begin{equation}
\def\arraystretch{1.75}
\begin{array}{c}
    \displaystyle
    \underset{\pm t}{S}^{k}(f(t_n, a_m, b_s, x^{n}_{m,s}, y^{n}_{m, s}))
        = f(t_{n \pm k}, a_m, b_s, x^{n \pm k}_{m,s}, y^{n \pm k}_{m, s}),
    \\
    \displaystyle
    \underset{\pm a}{S}^{k}(f(t_n, a_m, b_s, x^{n}_{m,s}, y^{n}_{m, s}))
        = f(t_{n}, a_{m \pm k}, b_s, x^{n}_{m \pm k,s}, y^{n}_{m \pm k, s}),
        \\
    \displaystyle
    \underset{\pm b}{S}^{k}(f(t_n, a_m, b_s, x^{n}_{m,s}, y^{n}_{m, s}))
        = f(t_{n}, a_{m}, b_{s \pm k}, x^{n}_{m,s \pm k}, y^{n}_{m, s \pm k}).
\end{array}
\end{equation}
The total differentiation operators are defined through the shifts as
\begin{equation}
\begin{array}{c}
\displaystyle
\dtp{D} = \frac{\dtp{S} - 1}{t_{n+1} - t_n},
\qquad
\dtm{D} = \frac{1 - \dtm{S}}{t_{n} - t_{n-1}},
\\
\\
\displaystyle
\dap{D} = \frac{\dap{S} - 1}{a_{m+1} - a_m},
\quad
\dam{D} = \frac{1 - \dam{S}}{a_{m} - a_{m-1}},
\qquad
\dbp{D} = \frac{\dbp{S} - 1}{b_{s+1} - b_s},
\quad
\dbm{D} = \frac{1 - \dbm{S}}{b_{s} - b_{s-1}},
\end{array}
\end{equation}
and the following notation is used for difference derivatives
\[
x_t = \dtp{D}(x),
\quad
\check{x}_t = \dtm{D}(x),
\quad
x_{t\check{t}} = \dtm{D}\dtp{D}(x),
\quad
x_a = \dap{D}(x),
\quad
\check{x}_a = \dam{D}(x),
\quad
x_{a\bar{a}} = \dam{D}\dap{D}(x),
\quad
\textrm{etc.}
\]

\subsection{Invariance of uniform orthogonal meshes}

Notice that all the shift and differentiation operators commute in any order on uniform orthogonal mesh
\begin{equation} \label{uniortMesh}
\hat{t} - t = t - \check{t} = \tau,
\qquad
a_+ - a = a - a_- = h^a,
\qquad
{}^+b - b = b - {}^-b = h^b,
\end{equation}
where $\tau >0$, $h^a>0$ and $h^b>0$ are small enough constant values.

\medskip

In the finite-difference case, in order to preserve uniform orthogonal meshes the
generator~$X = \xi^t \ppartial{t}+ \xi^a \ppartial{a}+ \xi^b \ppartial{b}$
must satisfy the following criteria~\cite{bk:Dorodnitsyn[1991],bk:Dorodnitsyn[2011]}
\begin{equation} \label{UniCrit}
\dtp{D}\dtm{D}(\xi^t) = 0,
\qquad
\dap{D}\dam{D}(\xi^a) = 0,
\qquad
\dbp{D}\dbm{D}(\xi^b) = 0,
\end{equation}
\begin{equation} \label{OrtCrit}
\underset{\pm b}{D}(\xi^a) = -\underset{\pm a}{D}(\xi^b),
\quad
\underset{\pm \tau}{D}(\xi^a) = -\underset{\pm a}{D}(\xi^t),
\quad
\underset{\pm \tau}{D}(\xi^b) = -\underset{\pm b}{D}(\xi^t),
\end{equation}

The transformation corresponding to~$X_{2}$ is related to the freedom of the
Lagrangian coordinates parametrization~\cite{bk:KaptsovMeleshko2019_2}
which should be restricted in the difference case due to~(\ref{UniCrit}) and~(\ref{OrtCrit}).

One can check that the generator~$X_{2}$ in its general form
which depend on an arbitrary function~$\psi$
does not satisfy~(\ref{UniCrit}),~(\ref{OrtCrit}).
As a particular example, consider
\[
\psi(a, b) = a^2 - b^2.
\]
From~(\ref{kernel})
it follows that $\xi^a = -2b$ and $\xi^b = -2a$ in this case, and
the orthogonality condition~(\ref{OrtCrit}) does not hold
\begin{equation}
\underset{+b}{D}(\xi^a) + \underset{+a}{D}(\xi^b)
= \underset{+b}{D}(-2b) + \underset{+a}{D}(-2a) = -4 \neq 0.
\end{equation}
For simplicity, we restrict our consideration to generators with coefficients of the form
\begin{equation}
\xi^t = 0,
\qquad
\xi^a = \alpha_1 a + \beta_1 b +\gamma_1,
\qquad
\xi^b = \alpha_2 a + \beta_2 b +\gamma_2,
\qquad
\alpha_i, \beta_i, \gamma_i = \textrm{const},
\quad
i = 1, 2,
\end{equation}
which satisfy the uniformness conditions~(\ref{UniCrit}).
Substituting $\xi^a$ and $\xi^b$ into the orthogonality condition~(\ref{OrtCrit}),
one obtains $\beta_1 = -\alpha_2$, i.e.,
\begin{equation}
\xi^a = \alpha_1 a -\alpha_2 b +\gamma_1,
\qquad
\xi^b = \alpha_2 a + \beta_2 b +\gamma_2.
\end{equation}
According to~(\ref{kernel}), one has the following restrictions on the function $\psi$
\begin{equation}
  \xi^a = \alpha_1 a - \alpha_2 b +\gamma_1 = \psi_b,
  \qquad
  \xi^b = \alpha_2 a + \beta_2 b +\gamma_2 = -\psi_a.
\end{equation}
Integrating the latter equations, one gets
\begin{equation}
\psi(a,b) = \alpha_1 a b - \frac{1}{2} \alpha_2 b^2 + \gamma_1 b + \chi_1(a)
= -\beta_2 a b - \frac{1}{2} \alpha_2 a^2 - \gamma_2 a + \chi_2(b),
\end{equation}
where $\chi_1$ and $\chi_2$ are some functions of their arguments.
Comparing the latter expressions for the function~$\psi$,
one obtains that $\beta_2 = -\alpha_1$, and
the function $\psi$ corresponding to the chosen particular solution is the following
\begin{equation} \label{PsiGenForm}
\psi(a, b) = \alpha_1 ab - \frac{\alpha_2}{2}(a^2 + b^2) - \gamma_2 a + \gamma_1 b + \delta,
\qquad
\delta = \textrm{const}.
\end{equation}
Substituting~(\ref{PsiGenForm}) into~(\ref{kernel}),
one obtains the following particular form of the generator $X_2$
\[
X_2^0 = (\alpha_1 a - \alpha_2 b + \gamma_1) \ppartial{a}
    + (\alpha_2 a - \alpha_1 b + \gamma_2) \ppartial{b},
\]
which results in the following set of shifting, inhomogeneous scaling and rotation generators
\begin{equation} \label{X20}
X_2^1 = \ppartial{a},
\qquad
X_2^2 = \ppartial{b},
\qquad
X_2^3 = a \ppartial{a} - b \ppartial{b},
\qquad
X_2^4 = b \ppartial{a} - a \ppartial{b}.
\end{equation}
The generators $X_2^3$ and $X_2^4$ form Lie algebras with the generators $X_1$, $X_2^1, X_2^2$, $Y_1$--$Y_8$,
but their commutator
\[
[X_2^3,X_2^4] = 2 b \ppartial{a} + 2 a \ppartial{b}
\]
does not satisfy orthogonality conditions~(\ref{OrtCrit}).
One has to choose either the generator~$X_2^3$ or the generator~$X_2^4$.
Further we prefer the generator~$X_2^3$,
and we restrict ourselves to the Lie algebra $X_1$, $X_2^1$---$X_2^3$, $Y_1$--$Y_8$.

\medskip

The generators $X_1$, $X_2^1$--$X_2^3$, $Y_1$--$Y_5$, $Y_7$ and $Y_8$ satisfy conditions~(\ref{UniCrit}) and~(\ref{OrtCrit}), and the uniform orthogonal mesh~(\ref{uniortMesh})
is invariant with respect to these generators.

\subsection{Jacobian invariance and mass conservation}

While construction of conservative schemes, it is important to preserve a finite-difference analogue
of the conservation of mass identity~(\ref{CLmass}).
Considering approximations for~(\ref{CLmass}) on the 8-point stencil\footnote{
Approximations on the chosen 8-point stencil can be then shifted to the left by the
operators $\dtm{D}$, $\dam{D}$ and~$\dbm{D}$.}
on a uniform orthogonal mesh in the form
\begin{multline}
\dtp{D}({\dbp{S}}^{i_1}(x_a) {\dap{S}}^{i_2}(y_b) - {\dap{S}}^{i_3}(x_b) {\dbp{S}}^{i_4}(y_a))
- \dap{D}({\dbp{S}}^{i_5}(x_t) {\dtp{S}}^{i_6}(y_b) - {\dtp{S}}^{i_7}(x_b) {\dbp{S}}^{i_8}(y_t)){}
\\
{}- \dbp{D}({\dtp{S}}^{i_9}(x_a) {\dap{S}}^{i_{10}}(y_t) - {\dap{S}}^{i_{11}}(x_t) {\dtp{S}}^{i_{12}}(y_a))= 0,
\end{multline}
where $i_1, ..., i_{12} \in \{0,1\}$ are unknown indices,
one states by direct computation that there are only two possible
difference analogues of equation~(\ref{CLmass}) that identically hold,
namely
\begin{equation} \label{approxMass}
\dtp{D}({}^+x_a y_b - x_b^+ y_a)
- \dap{D}({}^+x_t y_b - \hat{x}_b y_t)
- \dbp{D}(\hat{x}_a y_t - x_t^+ y_a) = 0,
\end{equation}
and
\begin{equation}
\dtp{D}(x_a y^+_b - {}^+y_a x_b)
- \dap{D}(x_t y_b - {}^+y_t \hat{x}_b)
- \dbp{D}(\hat{x}_a y_t^+ - x_t y_a) = 0.
\end{equation}

We consider linear combinations
\begin{equation} \label{jacApprox00}
\theta ({}^+x_a y_b - x_b^+ y_a) + (1 - \theta) (x_a y^+_b - {}^+y_a x_b),
\qquad
0 \leqslant \theta \leqslant 1
\end{equation}
as approximations for Jacobian~(\ref{jac}).
In order to choose the value of~$\theta$ we notice that Jacobian~(\ref{jac}) is invariant with respect to the generators $X_1$, $X_2$, $Y_1$--$Y_5$ (see Remark~\ref{rem:jac}).
Jacobian~(\ref{jac}) is a fundamental generating
differential invariant of the particle relabelling symmetry~\cite{bk:ClarksonBila[2006]},
and in the finite-difference case we would like to hold as much its geometric properties as possible.
Recall that in the previous sections we posed restrictions on the generator $X_2$ due to the mesh uniformness and orthogonality conditions, so the chosen particle relabelling symmetries of our interest are~$X_2^1$, $X_2^2$ and~$X_2^3$.
Applying the generators $X_1$, $X_2^1$--$X_2^3$ and $Y_1$--$Y_5$ to~(\ref{jacApprox00}),
one finds that the only value of~$\theta$ that preserves invariance with respect
to all the considered symmetries is~$\theta = 1/2$.
Thus, we prefer the following approximation for Jacobian~(\ref{jac})
\begin{equation} \label{JhApprox}
\dh{J}
= \frac{1}{2}\left(
    x_a y^+_b + {}^+x_a y_b
    - y_a x_b^+ - {}^+y_a x_b
\right).
\end{equation}

In this case, the conservation law of mass is the following identity
\begin{equation}
\dtp{D}({}^+x_a y_b + x_a y^+_b - x_b^+ y_a - {}^+y_a x_b)
- \dap{D}({}^+x_t y_b - \hat{x}_b y_t + x_t y_b - {}^+y_t \hat{x}_b)
- \dbp{D}(\hat{x}_a y_t - x_t^+ y_a + \hat{x}_a y_t^+ - x_t y_a) = 0,
\end{equation}
or
\begin{equation} \label{CLmassIdentity}
\dtp{D}({}^+x_a y_b + x_a y^+_b - x_b^+ y_a - {}^+y_a x_b)
- \dap{D}((x_t + {}^+x_t) y_b - \hat{x}_b (y_t + {}^+y_t ) \hat{x}_b)
- \dbp{D}(\hat{x}_a (y_t + y_t^+) - y_a (x_t + x_t^+)) = 0.
\end{equation}

\subsection{Approximations for the derivatives of the function $H(x,y)$}

Before proceeding any further, we consider the problem of approximating
the differential derivatives~$H_x$ and~$H_y$.
No finite-difference derivatives by the dependant variables~$x$ and~$y$ have been defined yet.
Generalizing the approach introduced by the authors in~\cite{bk:DorKapSW2020},
we notice that from the obvious differential relations
\begin{equation}
H_a = H_x x_a + H_y y_a,
\qquad
H_b = H_x x_b + H_y y_b,
\qquad
H_t = H_x x_t + H_y y_t
\end{equation}
it follows that
\begin{equation} \label{HxHy}
H_x = \frac{H_b y_t - H_t y_b}{x_b y_t - x_t y_b} = \frac{H_a y_t - H_t y_a}{x_a y_t - x_t y_a},
\qquad
H_y = \frac{H_t x_b - H_b x_t}{x_b y_t - x_t y_b} = \frac{H_t x_a - H_a x_t}{x_a y_t - x_t y_a}.
\end{equation}

Further we consider some approximation~$\Theta$ for the function~$H$.
For definiteness, we choose
\begin{equation}\label{Htheta}
\Theta = \frac{1}{2}(H+\check{H}).
\end{equation}

According to~(\ref{HxHy}), it seems natural to define approximations~$\Theta_x$
and~$\Theta_y$ for the continuous derivatives~$H_x$ and~$H_y$
through the known finite-difference
derivatives~$\Theta_t$ and~$\Theta_a$.
We choose the following approximations for the derivatives of~(\ref{Htheta})
\begin{equation} \label{ThetaXY}
\def\arraystretch{2.25}
\begin{array}{c}
\displaystyle
\Theta_x
= \frac{\Theta_a (y_t + \check{y}_t) - 2 \Theta_t y_a}{x_a (y_t + \check{y}_t) - (x_t + \check{x_t}) y_a}
= \frac{\Theta_b (y_t + \check{y}_t) - 2 \Theta_t y_b}{x_b (y_t + \check{y}_t) - (x_t + \check{x_t}) y_b},
\\
\displaystyle
\Theta_y
= \frac{2 \Theta_t x_a - \Theta_a (x_t + \check{x_t})}{x_a (y_t + \check{y}_t) - (x_t + \check{x_t}) y_a}
= \frac{2 \Theta_t x_b - \Theta_b (x_t + \check{x_t})}{x_b (y_t + \check{y}_t) - (x_t + \check{x_t}) y_b}.
\end{array}
\end{equation}
Notice that from the latter relations one derives that
\begin{equation} \label{ThetaTRel}
\dtp{D}(\Theta) = \Theta_t = \frac{x_t + \check{x_t}}{2} \Theta_x + \frac{y_t + \check{y_t}}{2}\Theta_y.
\end{equation}

Below we consider two approaches to the construction of invariant conservative schemes for the two-dimensional shallow water equations.

\bigskip

\subsection{Approach 1: Construction of two-dimensional schemes by extending the known one-dimensional scheme}

Recall that the case of an arbitrary bottom topography $H = H(x,y)$ is considered.

In case the bottom is arbitrary, the only generators that should be admitted by invariant schemes
are $X_1 = \ppartial{t}$ and $X_2^1$--$X_2^3$ which is given by~(\ref{X20}).
Such a poor set of generators makes the method of differential invariants~\cite{bk:Dorodnitsyn[2011]}
barely applicable as almost any arbitrary given discretization would be an invariant one.
As an alternative approach, it seems natural to choose schemes that
reduce to the invariant conservative scheme
\begin{equation} \label{simpleSchemeSW1}
x_{t\check{t}}
      + \dam{D}\left(
        \frac{1}{\hat{x}_a \check{x}_a}
      \right)
      - \frac{2}{x_t +\check{x}_t} \Theta_t
      = 0,
\end{equation}
\[
\tau  = \textrm{const},
\qquad
h^a = \textrm{const},
\]
previously constructed by the authors in~\cite{bk:DorKapSW2020}
for the one-dimensional shallow water equations~(\ref{scheme1d}).
Here~$\Theta$ is given by~(\ref{Htheta}).

There still a broad class of such schemes, and as an example one can choose the following one

\begin{equation} \label{approxF1}
F_1 = \dtm{D}(x_t)
+ \dam{D}\left(\frac{y_{\bar{b}}}{
    {}^-{\dh{\check{J}}}
    {}^-{\dh{\hat{J}}}
}\right)
- \dbm{D}\left(\frac{y_{\bar{a}}}{
    {\dh{\hat{J}}}^-
    {\dh{\check{J}}}^-
}\right) - \Theta_x = 0,
\end{equation}
\begin{equation} \label{approxF2}
F_2 = \dtm{D}(y_t)
- \dam{D}\left(\frac{x_{\bar{b}}}{
    {}^-{\dh{\check{J}}}
    {}^-{\dh{\hat{J}}}
}\right)
+ \dbm{D}\left(\frac{x_{\bar{a}}}{
    {\dh{\hat{J}}}^-
    {\dh{\check{J}}}^-
}\right) - \Theta_y = 0,
\end{equation}
where~$\dh{J}$ is given by equation~(\ref{JhApprox}),
and $\Theta_x$ and $\Theta_y$ are given by~(\ref{ThetaXY}).

\medskip

The sum of equations (\ref{approxF1}) and (\ref{approxF2}) is a inhomogenious~(see e.g.~\cite{bk:RozhdYanenko[1978]}) conservation law
\begin{equation}
\dtm{D}\left(x_t + y_t\right)
+\dam{D}\left(
    \frac{y_{\bar{b}} - x_{\bar{b}}}{
        {}^-{\dh{\check{J}}}
        {}^-{\dh{\hat{J}}}
    }
\right)
+\dbm{D}\left(
    \frac{x_{\bar{a}} - y_{\bar{a}}}{
        {\dh{\hat{J}}}^-
        {\dh{\check{J}}}^-
    }
\right)- \Theta_x - \Theta_y = 0
\end{equation}
which becomes a homogenious one in case $H = \textrm{const}$, namely
\begin{equation}
\dtm{D}\left(x_t + y_t\right)
+\dam{D}\left(
    \frac{y_{\bar{b}} - x_{\bar{b}}}{
        {}^-{\dh{\check{J}}}
        {}^-{\dh{\hat{J}}}
    }
\right)
+\dbm{D}\left(
    \frac{x_{\bar{a}} - y_{\bar{a}}}{
        {\dh{\hat{J}}}^-
        {\dh{\check{J}}}^-
    }
\right) = 0.
\end{equation}
In addition, in case $H = \textrm{const}$,
the following finite-difference analogue of the center of mass conservation~(\ref{CLcom})
is possessed by system~(\ref{approxF1}), (\ref{approxF2})
\begin{equation}
\dtm{D}\left( t (x_t + y_t) - x - y\right)
+\dam{D}\left(
    t\,\frac{y_{\bar{b}} - x_{\bar{b}}}{
        {}^-{\dh{\check{J}}}
        {}^-{\dh{\hat{J}}}
    }
\right)
+\dbm{D}\left(
    t\,\frac{x_{\bar{a}} - y_{\bar{a}}}{
        {\dh{\hat{J}}}^-
        {\dh{\check{J}}}^-
    }
\right) = 0.
\end{equation}

\subsection{Approach 2: Direct algebraic construction of conservative schemes}

An alternative approach is to consider the conservation law of energy in some divergent form, e.g.
\begin{equation} \label{CLenergyDIV}
\dtp{D}\left(
    \frac{\check{x}_t^2 + \check{y}_t^2}{2} + {J}_0^{-1}
    - \Theta
\right)
+\dap{D}\left(
    A \dam{S}( N^A (J_0 \hat{J}_0)^{-1} )
\right)
+\dbp{D}\left(
    B \dbm{S} ( N^B (J_0 \hat{J}_0)^{-1} )
\right) = 0,
\end{equation}
where
\begin{equation} \label{ABapprox}
    A = \frac{x_t + \check{x}_t}{2} \tilde{y}_b - \frac{y_t + \check{y}_t}{2} \tilde{x}_b,
    \qquad
    B = -\frac{x_t + \check{x}_t}{2} \tilde{y}_a + \frac{y_t + \check{y}_t}{2} \tilde{x}_a,
\end{equation}
$\tilde{x}_a$, $\tilde{x}_b$, $\tilde{y}_a$, $\tilde{y}_b$ are some approximations for the corresponding partial derivatives,
$J_0$ is some approximation for Jacobian~(\ref{jac}).

In order to approximate the conservation law of energy~(\ref{CLenergy})
there should be $N^A \to 1$ and $N^B \to 1$.
The exact form of the finite-difference terms $N^A$ and $N^B$ will be stated later.

Notice that there is a reason for choosing approximations precisely of the form~(\ref{ABapprox}).
It was mentioned by the authors~\cite{bk:DorKapSW2020,bk:ChevDorKap2020}
that conservation law multipliers corresponding to energy conservation laws
of schemes for one-dimensional equations often have the form
$\frac{1}{2}(x_t + \check{x}_t)$, and, thus, in the two-dimensional case
we assume them to be
\begin{equation} \label{lambdas}
\frac{x_t + \check{x}_t}{2}
\qquad
\text{and}
\qquad
\frac{y_t + \check{y}_t}{2}.
\end{equation}
As it is shown below, by means of the finite-difference Leibniz rule~(see e.g.~\cite{bk:Dorodnitsyn[2011]})
equation~(\ref{CLenergyDIV}) can be brought in such a form where the terms~(\ref{lambdas})
of the approximations~(\ref{ABapprox}) become conservation law multipliers.

\medskip

By applying the finite-difference Leibniz rule, one rewrites~(\ref{CLenergyDIV})
in the following form
\begin{multline}
  \dtp{D}\left(
    \frac{\check{x}_t^2 + \check{y}_t^2}{2} + {J}_0^{-1}
    - \Theta
\right)
+ \left(
        \frac{x_t + \check{x}_t}{2} \tilde{y}_b
        - \frac{y_t + \check{y}_t}{2} \tilde{x}_b
    \right) \dam{D}(N^A (J_0 \hat{J}_0)^{-1})
    + (J_0 \hat{J}_0)^{-1} N^A \dap{D}(A)
\\
+ \left(
        -\frac{x_t + \check{x}_t}{2} \tilde{y}_a
        + \frac{y_t + \check{y}_t}{2} \tilde{x}_a
    \right) \dbm{D}(N^B (J_0 \hat{J}_0)^{-1})
    + (J_0 \hat{J}_0)^{-1} N^B \dbp{D}(B) = 0.
\end{multline}
Expanding the first term $\dtp{D}(\cdots)$ of the latter equation, one gets
\begin{multline}
\frac{({x}_t + \check{x}_t)({x}_t - \check{x}_t)}{2\tau}
+ \frac{({y}_t + \check{y}_t)({y}_t - \check{y}_t)}{2\tau}
- \dtp{D}(J_0) (J_0 \hat{J}_0)^{-1}
- \Theta_t
\\
+ \left(
        \frac{x_t + \check{x}_t}{2} \tilde{y}_b
        - \frac{y_t + \check{y}_t}{2} \tilde{x}_b
    \right) \dam{D}(N^A (J_0 \hat{J}_0)^{-1})
    + (J_0 \hat{J}_0)^{-1} N^A \dap{D}(A)
\\
+ \left(
        -\frac{x_t + \check{x}_t}{2} \tilde{y}_a
        + \frac{y_t + \check{y}_t}{2} \tilde{x}_a
    \right) \dbm{D}(N^B (J_0 \hat{J}_0)^{-1})
    + (J_0 \hat{J}_0)^{-1} N^B \dbp{D}(B) = 0.
\end{multline}
Taking~(\ref{ThetaTRel}) into account
and collecting the terms with respect to $(x_t + \check{x_t})$ and $(y_t + \check{y_t})$, one derives
the following equation
\begin{multline} \label{FDCLenergyDID}
\frac{x_t + \check{x}_t}{2} \left[
    x_{t\check{t}} + \tilde{y}_b \dam{D}(N^A (J_0 \hat{J}_0)^{-1})
    - \tilde{y}_a \dbm{D}(N^B (J_0 \hat{J}_0)^{-1})
    - \Theta_x
\right]
\\
+  \frac{y_t + \check{y}_t}{2} \left[
    y_{t\check{t}} - \tilde{x}_b \dam{D}(N^A (J_0 \hat{J}_0)^{-1})
    + \tilde{x}_a \dbm{D}(N^B (J_0 \hat{J}_0)^{-1})
    -\Theta_y
\right]
\\
+ \left(\dtp{D}{J_0} - N^A \dap{D}A - N^B \dbp{D}B\right)(J_0 \hat{J}_0)^{-1} = 0.
\end{multline}
To eliminate the term
\begin{equation}\label{NondivTerm}
\dtp{D}{J_0} - N^A \dap{D}A - N^B \dbp{D}B,
\end{equation}
one can put
\begin{equation} \label{Nlimits}
N^A = \frac{\dap{D} A_0}{\dap{D} A} \to 1,
\qquad
N^B = \frac{\dap{D} B_0}{\dap{D} B} \to 1,
\end{equation}
where $A_0 \neq 0$ and $B_0 \neq 0$ are some approximations alternative to $A$ and $B$.
Thus, one brings~(\ref{NondivTerm}) to the divergent form
\begin{equation}
\dtp{D}{J_0} - \dap{D}A_0 - \dbp{D}B_0.
\end{equation}
The latter expression approximates the conservation law of mass and vanishes in the continuous limit.
Thus, the final step is to choose some specific approximations $J_0$, $A_0$ and $B_0$
to make \emph{identically} hold the equation
\begin{equation}
\dtp{D}{J_0} - \dap{D}A_0 - \dbp{D}B_0 = 0.
\end{equation}
We choose, for example, the invariant approximation~(\ref{JhApprox}) as $J_0$,
and, according to~(\ref{CLmassIdentity}), the following approximations as $A_0$ and $B_0$
\begin{equation}
A_0 = \frac{1}{2}\left((x_t + {}^+x_t) y_b - (y_t + {}^+y_t) \hat{x}_b\right),
\qquad
B_0 = \frac{1}{2}\left((y_t + y_t^+) \hat{x}_a - (x_t + x_t^+ ) y_a \right).
\end{equation}

Substituting the latter approximations into~(\ref{FDCLenergyDID}), one obtains the scheme
\begin{equation} \label{ConsrvScheme}
\def\arraystretch{2}
\begin{array}{c}
\displaystyle
F_1 = x_{t\check{t}}
+ \tilde{y}_b \dam{D}\left[
    \frac{1}{J_0 \hat{J}_0}
    \frac {\dap{D} ((x_t + {}^+x_t) y_b - (y_t + {}^+y_t) \hat{x}_b)}
            {\dap{D} ((x_t + \check{x}_t) \tilde{y}_b - (y_t + \check{y}_t) \tilde{x}_b)}
     \right]
     \\
     \displaystyle
    - \tilde{y}_a \dbm{D}\left[
        \frac{1}{J_0 \hat{J}_0}
        \frac {\dbp{D} ((y_t + y_t^+) \hat{x}_a - (x_t + x_t^+ ) y_a)}
            {\dbp{D} ((y_t + \check{y}_t) \tilde{x}_a - (x_t + \check{x}_t) \tilde{y}_a)}
    \right]
- \Theta_x
= 0,
\\
\displaystyle
F_2 = y_{t\check{t}}
- \tilde{x}_b \dam{D}\left[
    \frac{1}{J_0 \hat{J}_0}
    \frac {\dap{D} ((x_t + {}^+x_t) y_b - (y_t + {}^+y_t) \hat{x}_b)}
            {\dap{D} ((x_t + \check{x}_t) \tilde{y}_b - (y_t + \check{y}_t) \tilde{x}_b)}
     \right]
     \\
     \displaystyle
    + \tilde{x}_a \dbm{D}\left[
        \frac{1}{J_0 \hat{J}_0}
        \frac {\dbp{D} ((y_t + y_t^+) \hat{x}_a - (x_t + x_t^+ ) y_a)}
            {\dbp{D} ((y_t + \check{y}_t) \tilde{x}_a - (x_t + \check{x}_t) \tilde{y}_a)}
    \right]
-\Theta_y
= 0,
\end{array}
\end{equation}
on uniform orthogonal mesh~(\ref{uniortMesh}),
where
\begin{equation} \label{jac0}
J_0 = \underset{h}{J} = \frac{1}{2}\left(
    x_a y^+_b + {}^+x_a y_b
    - y_a x_b^+ - {}^+y_a x_b
\right),
\end{equation}
and $\Theta_x$, $\Theta_y$ are given by equations~(\ref{ThetaXY}).


\smallskip

The conservation law of energy for scheme~(\ref{ConsrvScheme}) is
\begin{multline} \label{CLenergy2D}
\frac{x_t + \check{x}_t}{2} F_1
+ \frac{y_t + \check{y}_t}{2} F_2
=
\dtp{D}\left(
    \frac{\check{x}_t^2 + \check{y}_t^2}{2} + J_0^{-1}
    - \Theta
\right)
\\
+\frac{1}{2}\dap{D}\left(
    ((x_t + \check{x}_t) \tilde{y}_b - (y_t + \check{y}_t) \tilde{x}_b)
    \frac {\dam{D} ((x_t + {}^+x_t) y_b - (y_t + {}^+y_t) \hat{x}_b)}
        {\dam{D} ((x_t + \check{x}_t) \tilde{y}_b - (y_t + \check{y}_t) \tilde{x}_b)}
    \dam{S}((J_0 \hat{J}_0)^{-1})
\right)
\\
+\frac{1}{2}\dbp{D}\left(
    ((y_t + \check{y}_t) \tilde{x}_a - (x_t + \check{x}_t) \tilde{y}_a)
    \frac {\dbm{D} ((y_t + y_t^+) \hat{x}_a - (x_t + x_t^+ ) y_a)}
            {\dbm{D} ((y_t + \check{y}_t) \tilde{x}_a - (x_t + \check{x}_t) \tilde{y}_a)}
    \dbm{S} ((J_0 \hat{J}_0)^{-1})
\right) = 0,
\end{multline}
and the conservation law of mass is just an identity
\begin{equation} \label{CLmass2D}
\displaystyle
\dtp{D}{J_0}
- \frac{1}{2} \dap{D}((x_t + {}^+x_t) y_b - (y_t + {}^+y_t) \hat{x}_b)
- \frac{1}{2} \dbp{D} ((y_t + y_t^+) \hat{x}_a - (x_t + x_t^+ ) y_a) = 0.
\end{equation}
Thus, we have the conservation laws of mass and energy by construction.

\subsection{Reductions of scheme~(\ref{ConsrvScheme}) to the one-dimensional case}

In the present section we consider one-dimensional reductions for scheme~(\ref{ConsrvScheme})
and its modifications.

\medskip

In the one-dimensional case, Jacobian~(\ref{jac0}) becomes $J_0 = x_a$.
From equations~(\ref{ThetaXY}) and the condition~$H_y = 0$ (or $\Theta_y = 0$)
it follows that
\begin{equation}
\Theta_t = \frac{x_t + \check{x}_t}{2} \frac{\Theta_a}{x_a}.
\end{equation}
Taking the latter and conditions~(\ref{Nlimits}) into account, one reduces scheme~(\ref{ConsrvScheme}) to the form
\begin{equation} \label{reduct1}
F = x_{t\check{t}} + \dam{D}\left( \frac{2 x_{ta}}{x_a \hat{x}_a (x_{ta} + \check{x}_{ta})} \right) - \Theta_x = 0.
\end{equation}
The conservation law of energy~(\ref{CLenergy2D}) possesses the following form
\begin{equation}
\frac{x_t + \check{x}_t}{2} F = \dtp{D}\left(
    \frac{\check{x}_t^2}{2}
    + \frac{1}{x_a}
    - \Theta
\right)
+ \dap{D}\left(
    (x_t + \check{x}_t) \frac{x_{t{\bar{a}}}}{x_{\bar{a}} \hat{x}_{\bar{a}} (x_{t\bar{a}} + \check{x}_{t\bar{a}})}
\right) = 0,
\end{equation}
and the conservation law of mass~(\ref{CLmass2D}) is
\begin{equation}
\dtp{D}(x_a) - \dap{D}(x_t) = 0.
\end{equation}

Notice that the reduction~(\ref{reduct1}) depend on the third difference derivatives $x_{ta\bar{a}}$ and $\check{x}_{ta\bar{a}}$. The one-dimensional schemes constructed by the authors in~\cite{bk:Dorodnitsyn[2011]} are defined on 9-point finite-difference stencil, and the higher difference derivatives they depend on are of the second order. To obtain a better reduction, we modify scheme~(\ref{ConsrvScheme}) as follows.

By shifting Jacobian~(\ref{jac0}) along the time axis, we consider its modified version on the extended stencil
\begin{equation}
J_1 = \frac{1}{2}\left(
    {}^+x_a \check{y}_b +  \check{x}_a y^+_b
    - \check{y}_a x_b^+ - {}^+y_a \check{x}_b
\right).
\end{equation}
One can check that $J_1$ admits all the generators $X_1$, $X_2^1$--$X_2^3$ and $Y_1$--$Y_5$.

According to the latter changes, the conservation law of mass~(\ref{CLmass2D}) becomes
\begin{equation}
\displaystyle
\dtp{D}{J_1}
- \frac{1}{2} \dap{D}(\check{x}_t y_b + {}^+x_t \check{y}_b - \check{y}_t \hat{x}_b - {}^+y_t x_b)
- \frac{1}{2} \dbp{D} (\check{y}_t \hat{x}_a + y_t^+ x_a - \check{x}_t y_a -  x_t^+ \check{y}_a) = 0,
\end{equation}
and scheme~(\ref{ConsrvScheme}) possesses the form
\begin{equation}
\def\arraystretch{2}
\begin{array}{c}
\displaystyle
F_1 = x_{t\check{t}}
+ \tilde{y}_b \dam{D}\left[
    \frac{1}{J_1 \hat{J}_1}
    \frac {\dap{D} (\check{x}_t y_b + {}^+x_t \check{y}_b - \check{y}_t \hat{x}_b - {}^+y_t x_b)}
            {\dap{D} ((x_t + \check{x}_t) \tilde{y}_b - (y_t + \check{y}_t) \tilde{x}_b)}
     \right]
     \\
     \displaystyle
    - \tilde{y}_a \dbm{D}\left[
        \frac{1}{J_1 \hat{J}_1}
        \frac {\dbp{D} (\check{y}_t \hat{x}_a + y_t^+ x_a - \check{x}_t y_a -  x_t^+ \check{y}_a)}
            {\dbp{D} ((y_t + \check{y}_t) \tilde{x}_a - (x_t + \check{x}_t) \tilde{y}_a)}
    \right]
- \Theta_x
= 0,
\\
\displaystyle
F_2 = y_{t\check{t}}
- \tilde{x}_b \dam{D}\left[
    \frac{1}{J_1 \hat{J}_1}
    \frac {\dap{D} (\check{x}_t y_b + {}^+x_t \check{y}_b - \check{y}_t \hat{x}_b - {}^+y_t x_b)}
            {\dap{D} ((x_t + \check{x}_t) \tilde{y}_b - (y_t + \check{y}_t) \tilde{x}_b)}
     \right]
     \\
     \displaystyle
    + \tilde{x}_a \dbm{D}\left[
        \frac{1}{J_1 \hat{J}_1}
        \frac {\dbp{D} (\check{y}_t \hat{x}_a + y_t^+ x_a - \check{x}_t y_a -  x_t^+ \check{y}_a)}
            {\dbp{D} ((y_t + \check{y}_t) \tilde{x}_a - (x_t + \check{x}_t) \tilde{y}_a)}
    \right]
-\Theta_y
= 0.
\end{array}
\end{equation}
The conservation law of energy~(\ref{CLenergy2D}) becomes
\begin{multline}
\dtp{D}\left(
    \frac{\check{x}_t^2 + \check{y}_t^2}{2} + J_1^{-1}
    - \Theta
\right)
\\
+\frac{1}{2}\dap{D}\left(
    ((x_t + \check{x}_t) \tilde{y}_b - (y_t + \check{y}_t) \tilde{x}_b)
    \frac {\dam{D} (\check{x}_t y_b + {}^+x_t \check{y}_b - \check{y}_t \hat{x}_b - {}^+y_t x_b)}
        {\dam{D} ((x_t + \check{x}_t) \tilde{y}_b - (y_t + \check{y}_t) \tilde{x}_b)}
    \dam{S}((J_1 \hat{J}_1)^{-1})
\right)
\\
+\frac{1}{2}\dbp{D}\left(
    ((y_t + \check{y}_t) \tilde{x}_a - (x_t + \check{x}_t) \tilde{y}_a)
    \frac {\dbm{D} (\check{y}_t \hat{x}_a + y_t^+ x_a - \check{x}_t y_a -  x_t^+ \check{y}_a)}
            {\dbm{D} ((y_t + \check{y}_t) \tilde{x}_a - (x_t + \check{x}_t) \tilde{y}_a)}
    \dbm{S} ((J_1 \hat{J}_1)^{-1})
\right) = 0.
\end{multline}

Finally, the reduced one-dimensional scheme is
\begin{equation} \label{reduct2}
F = x_{t\check{t}} + \dam{D}\left( \frac{4}{(x_a + \check{x}_a) (x_a + \hat{x}_a)} \right) - \Theta_x = 0,
\end{equation}
\[
h^a = \textrm{const},
\qquad
\tau = \textrm{const},
\]
and the corresponding conservation laws of mass and energy have the following forms
\begin{equation}
\dtp{D}\left( x_a + \check{x}_a \right) - \dap{D}\left( x_t + \check{x}_t \right)= 0,
\end{equation}
\begin{equation}
\dtp{D}\left(
    \frac{\check{x}_t^2}{2}
    + \frac{2}{x_a + \check{x}_a}
    - \Theta
\right)
+ \dap{D}\left(
    \frac{2 (x_t + \check{x}_t)}{(x_{\bar{a}} + \check{x}_{\bar{a}}) (x_{\bar{a}} + \hat{x}_{\bar{a}})}
\right) = 0.
\end{equation}

In~\cite{bk:Dorodnitsyn[2011]}, with the help of the finite-difference analogue of the direct method~\cite{bk:BlumanAnco_adjoint[1996]},
the authors have obtained a family of invariant conservative schemes for the one-dimensional shallow water equations.
As a simplest example of such a scheme, the authors considered scheme~(\ref{simpleSchemeSW1}). One can check that scheme~(\ref{reduct2}) is found among the obtained family of the one-dimensional conservative schemes as well.


\section{Conclusion}

The group classification of the two-dimensional shallow water equations
with variable bottom topography~$H=p x^{2}+2cxy+by^{2}+q_{1}x+q_{2}y+q_{0}$ in mass Lagrangian
coordinates is performed.
The advantage of studying the shallow water
equations in Lagrangian coordinates is that in Lagrangian coordinates
they have a variational structure. This variational representation
allows one to apply Noether's theorem for constructing conservation
laws.
The classification results of the considered case are presented in Table~\ref{tab:classifQuadric}
and formulated in Theorem~\ref{th:th1}.
If the function~$H(x,y)$ is either of the form~$H(x,y)= p (x^{2}+y^{2})$
(corresponding to a circular paraboloid bottom) or $H(x,y)=\textrm{const}$ (corresponding to a plane bottom),
then system of the shallow water equations~(\ref{SWeulr1})--(\ref{SWeulr3})
can be reduced to the gas dynamics equations of a polytropic gas with the exponent~$\gamma=2$~\cite{bk:Meleshko2020}.

\smallskip

Notice that the admitted Lie algebra of the original two-dimensional shallow water equations
contains infinite algebra of relabelling operators. We restrict this algebra to preserve
a difference mesh orthogonality and uniformness, and keep all the rest symmetry of the equations.
New invariant schemes for the two-dimensional shallow water equations with arbitrary
bottom topography $H(x,y)$ in Lagrangian coordinates on uniform orthogonal meshes are proposed.
The schemes are constructed either by extending the known one-dimensional schemes
or by direct algebraic construction.
As it was mentioned, in case the bottom is arbitrary, such schemes can be constructed on uniform orthogonal meshes.
At the same time, there is a rather complicated problem of
approximation of the derivatives~$H_x$ and~$H_y$ of the arbitrary bottom~$H(x,y)$
for which there are no obvious representations of difference differentiation operators.
This problem is discussed in a separate section of the paper.

Among the proposed schemes there are schemes possessing the conservation laws of mass and energy.
In case of a horizontal bottom~$H(x, y) = \textrm{const}$, some of the schemes have conservative form and possess
conservation laws of momentum and the center-of-mass law.

Finally, it is shown that the proposed schemes can be reduced to
the known one-dimensional schemes previously constructed by the authors in~\cite{bk:DorKapSW2020}.

\section*{Acknowledgements}

The research was supported by Russian Science Foundation Grant No 18-11-00238
``Hydro\-dyna\-mics-type equations: symmetries, conservation
laws, invariant difference schemes''.
E.K.~acknowledges Suranaree University of Technology for
Full-time Master Researcher Fellowship.


\end{document}